\documentclass[11pt, onecolumn]{article}

\usepackage{fancyhdr}
\usepackage[english]{babel}
\usepackage{abstract}

\usepackage{amsmath}
\usepackage{amssymb}
\usepackage{euscript}
\usepackage{hyperref}

\usepackage{cyr}
\usepackage{epsfig}

\makeatletter
\renewcommand{\@oddhead}{}
\renewcommand{\@oddfoot}{\hfill ---~\thepage~---\hfill}
\makeatother

\fancypagestyle{firststyle} 
{
\fancyhead[L]{\small Published in Astronomy Letters, 2013, Vol. 39, No.10, pp. 717-728 \\ original russian text: Pis'ma v Astronomicheskii Zhurnal, 2013, Vol. 39, No. 10, pp. 797-809.\hfill}
\fancyfoot[L]{\hfill ---~\thepage~---\hfill}
 
}

\begin{document}
\thispagestyle{firststyle}

\begin{center}
\Large H$_{2}$O Maser Emission in Circumstellar Envelopes around AGB Stars: Physical Conditions in Gas-Dust Clouds

\vspace{0.5cm}
\large A.V. Nesterenok
\vspace{0.5cm}

\normalsize Ioffe Physical-Technical Institute, Polytechnicheskaya St. 26, Saint~Petersburg, 194021 Russia

e-mail: alex-n10@yandex.ru
\end{center}

 \begin{abstract} 
  \noindent
The pumping of 22.2-GHz H$_2$O masers in the circumstellar envelopes of asymptotic giant branch stars has been simulated numerically. The physical parameters adopted in the calculations correspond to those of the circumstellar envelope around IK Tau. The one-dimensional plane-parallel structure of the gas-dust cloud is considered. The statistical equilibrium equations for the H$_2$O level populations and the thermal balance equations for the gas-dust cloud are solved self-consistently. The calculations take into account 410 rotational levels belonging to the five lowest vibrational levels of H$_2$O. The stellar radiation field is shown to play an important role in the thermal balance of the gas-dust cloud due to the absorption of emission in rotational-vibrational H$_2$O lines. The dependence of the gain in the 22.2-GHz maser line on the gas density and H$_2$O number density in the gas-dust cloud is investigated. Gas densities close to the mean density of the stellar wind, 10$^7$-10$^8$ cm$^{-3}$, and a high relative H$_2$O abundance, more than 10$^{-4}$, have been found to be the most likely physical conditions in maser sources.
\end{abstract}

\smallskip

DOI: 10.1134/S106377371309003X

\noindent

\section*{\textmd{Introduction}}

The 22.2-GHz H$_2$O maser line corresponds to the permitted electric dipole transition between the rotational levels of the ground vibrational level of ortho-H$_2$O $J_{K_{a}K_{c}}$ = $6_{16} \to 5_{23}$, where $K_a$ and $K_c$ are the asymptotic quantum numbers that characterize the projections of the angular momentum vector onto the molecule's internal axes. H$_2$O maser emission is observed in many astrophysical objects: in star-forming regions, in accretion disks around compact massive objects, and in the expanding circumstellar envelopes of late-type stars. Owing to their high angular resolution and sensitivity, present-day radio interferometers allow the spatial distribution of maser sources to be mapped in detail. Numerical simulations of the maser pumping process and comparison of calculations with observational data allow the physical conditions in maser sources to be determined.

The 22.2-GHz H$_2$O maser emission in the circumstellar envelopes of asymptotic giant branch (AGB) stars is produced in clouds of gas and dust that are located at distances of about 10-60 AU from the star and have sizes of 2-4 AU (Bains et al. 2003). In the circumstellar envelopes of red supergiants, the distances and sizes of the maser sources are larger by several times (Richards et al. 1998, 1999; Murakawa et al. 2003). A correlation is observed between the sizes of the maser sources and the radius of the parent star. The filling factor of the circumstellar envelope by maser sources does not exceed 0.01 (Richards et al. 2012). The H$_2$O maser emission in the circumstellar envelopes of late-type stars is produced in compact gas-dust clouds.

Collisional pumping is considered as the main H$_2$O maser pumping mechanism (Yates et al. 1997). The pumping of H$_2$O masers in the circumstellar envelopes of late-type stars was numerically simulated by Deguchi (1977), Cooke and Elitzur (1985), Humphreys et al. (2001), Babkovskaia and Poutanen (2006), and other authors. The rotational levels belonging to the ground vibrational level and the first excited vibrational level of the molecule were taken into account by these authors. They used a method of level population calculations based on the photon escape probability. Note that Babkovskaia and Poutanen (2006) consider a self-consistent H$_2$O maser model by taking into account the main gas-dust cloud heating and cooling processes.

The goal of this paper is to investigate the physical conditions necessary for the generation of intense H$_2$O maser emission in gas-dust clouds in the circumstellar envelopes of AGB stars. The statistical equilibrium equations for the H$_2$O level populations and the thermal balance equations for gas and dust in the cloud are solved self-consistently. The accelerated $\Lambda$-iteration method is used in level population calculations (Rybicki and Hummer 1991). The calculations take into account 410 rotational levels belonging to the five lowest vibrational levels of H$_2$O. The influence of the radiation field of the parent star on the thermal balance of gas in the gas-dust cloud is investigated.

\noindent

\section*{\textmd{Model Parameters}}
\noindent
The maser sources generally have a complex spatial structure and consist of a multitude of bright spots, each of which is characterized by its relative velocity. The maser intensity depends significantly on the geometry of the gas-dust cloud and the gas velocity field in the cloud. In the expanding envelopes of late-type stars, the clouds of gas and dust probably have a flattened shape due to a difference between the cloud expansion velocities in the radial and tangential directions (Alcock and Ross 1986). The shock induced by stellar pulsations can also lead to a flat cloud shape. We consider the one-dimensional model of a flat gas-dust cloud (see Fig. 1).

\begin{figure}[t]
	\centering
	\includegraphics[width = 0.6\textwidth]{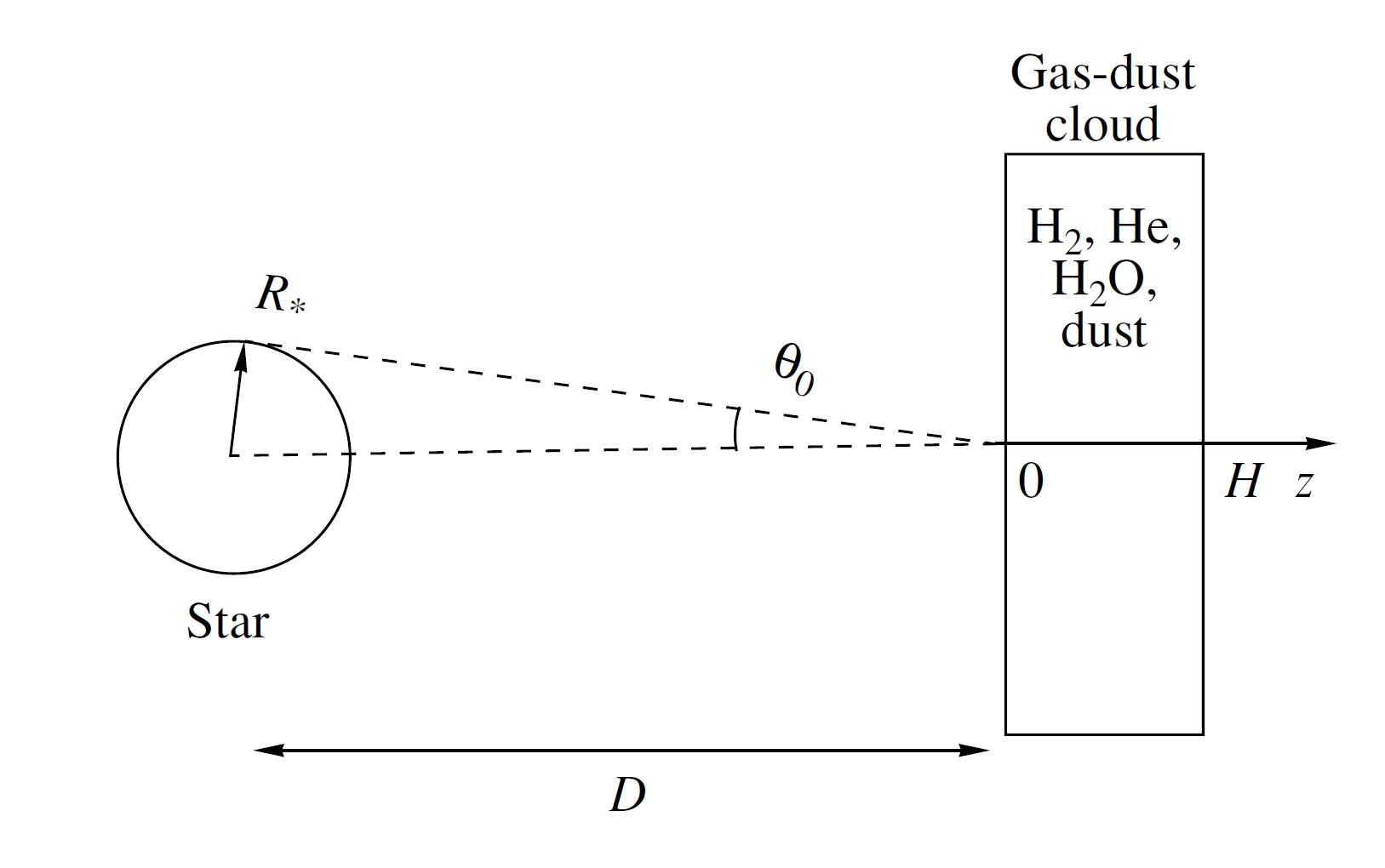}
		\caption{\small{Gas-dust cloud model.}}
		\label{fig1}
\end{figure}

The star's physical parameters adopted in our calculations correspond to those of the AGB star IK Tau (Monnier et al. 2004). IK Tau is an M-type star and is at a distance of about 250 pc (Olofsson et al. 1998). The estimates of the mass loss rate for a red giant $\dot M$ lie within the range from $4 \times 10^{-6}$ (Neri et al. 1998) to $3 \times 10^{-5}$ $M_\odot$ yr$^{-1}$ (Gonzalez Delgado et al. 2003). In our model, the geometrical sizes of the gas-dust cloud and the distance from the cloud to the star correspond to the mean values of these parameters for the maser sources in the envelope of IK Tau (Bains et al. 2003; Richards et al. 2011, 2012). The turbulent velocity in the gas-dust cloud is taken to be 1.5 km s$^{-1}$ (Decin 2012). The gas velocity gradient in the cloud is assumed to be 0.22 km s$^{-1}$ AU$^{-1}$ --- the mean stellar wind velocity gradient in the inner region of the gas-dust envelope around IK Tau, where the H$_2$O maser emission is observed (Richards et al. 2012). In the inner regions of the gas-dust envelopes around cool late-type stars, hydrogen is contained mainly in the form of H$_2$ molecules (Glassgold and Huggins 1983). The abundance of helium atoms relative to the total number density of hydrogen nuclei is assumed to be $N_{He}/2N_{H_2} = 0.1$. The relative abundance of H$_2$O molecules in the stellar wind depends on the carbon-to-oxygen ratio in the stellar atmosphere and can vary in a wide range (Cherchneff 2006). The ortho-to-para-H$_2$O ratio in our calculations is taken to be 3, which corresponds to the thermodynamic equilibrium at temperatures $>50$ K (Decin et al. 2010). The physical parameters of the model adopted in our calculations are given in the table.

The mean molecular hydrogen number density in the stellar wind at distance $D$ from the stellar center is estimated to be

\begin{equation}
\displaystyle
N_{H_2} = 0.71 \times \frac{\dot M}{4 \pi D^2 v m_{H_2}} = 5.4 \times 10^7 \left ( \frac{\dot M}{10^{-5} M_\odot \text{yr}^{-1}} \right) \left ( \frac{v}{10 \text{ km s}^{-1}} \right) ^{-1} \left ( \frac{D}{30 \text{ AU}} \right)^{-2} \text{cm}^{-3},
\end{equation}

\noindent
where the coefficient 0.71 corresponds to the hydrogen mass fraction in the stellar wind, $m_{H_2}$ is the mass of the hydrogen molecule, and $v$ is the expansion velocity of the gas-dust envelope at distance $D$.

~\\
~\\
\begin{tabular}{l@{\quad\quad}|@{\quad\quad}l}
\multicolumn{2}{l}{\large\bf Table} \\ [5pt]
\hline \\ [-2ex]
Stellar radius & $R_*$ = 2.5 AU \\ [3pt]
Stellar surface temperature & $T_*$ = 2300 K \\ [3pt]
Distance from stellar center to cloud & $D$ = 30 AU \\ [3pt]
Cloud thickness & $H$ = 3 AU \\ [3pt]
Turbulent velocity & $v_{\text{turb}}$ = 1.5 km s$^{-1}$ \\ [3pt]
Gas velocity gradient in cloud & $K_V$ = 0.22 km s$^{-1}$ AU$^{-1}$ \\ [3pt]
Range of H$_2$ number densities & $10^7$ cm$^{-3} < N_{H_2} < 5 \times 10^9$ cm$^{-3}$ \\ [3pt]
Range of H$_2$O number densities & $10^3$ cm$^{-3} < N_{H_2O} < 5 \times 10^5$ cm$^{-3}$ \\ [3pt]
Ortho-to-para-H$_2$O ratio & $N_{\text{o-}H_2O}/N_{\text{p-}H_2O} = 3$ \\[3pt] \hline
\end{tabular}
~\\
~\\

\section*{\textmd{Calculating the Level Populations of H$_2$O Molecules}}

\subsection*{\textmd{\textit{The Radiative Transfer Equation in Molecular Lines}}}

Consider a gas-dust cloud that consists of a mixture of H$_2$ and H$_2$O molecules, He atoms, and dust particles and is in the radiation field of its parent star (see Fig. 1). The cloud sizes along two coordinate axes are much larger than those along the third $z$ coordinate axis. We assume that there is a gas velocity gradient along the z axis. For simplification, the physical parameters of the cloud (the gas and dust temperatures, the number densities of atoms and molecules, and the dust content) are assumed to be independent of the coordinates. However, the H$_2$O level populations are considered as functions of the $z$ coordinate.

In the one-dimensional geometry, the intensity of radiation $I$ at frequency $\nu$ depends on depth $z$ and angle $\theta$ between the $z$ axis and the radiation direction. The quantity $\mu = cos \theta$ is used instead of the variable angle $\theta$. The point $z = 0$ corresponds to the cloud boundary facing the star. The radiative transfer equation can be written as

\begin{equation}
\mu \frac{dI(z,\mu,\nu)}{dz} = -\kappa(z,\mu,\nu) I(z,\mu,\nu) + \varepsilon(z,\mu,\nu),
\end{equation}

\noindent
where $I(z,\mu,\nu)$ is the intensity of radiation at frequency $\nu$ in direction $\mu$, $\varepsilon(z,\mu,\nu)$ is the emission coefficient, and $\kappa(z,\mu,\nu)$ is the absorption coefficient. The boundary condition for Eq. (2) at $z = 0$ is $I(0,\mu,\nu) = I_0(\mu,\nu)$, $\mu > 0$, where $I_0(\mu,\nu)$ is the intensity of the external radiation field. At $z = H$ we will take the boundary condition to be $I(H,\mu,\nu) = 0$, $\mu < 0$. The radiation from the parent star is considered as the external radiation field. The following relation holds for the intensity of the external radiation field:

\begin{equation}
\displaystyle
I_0(\mu, \nu) = \left\{
\begin{array}{ll}
B(T_*, \nu), & \mu_0 \leq \mu \leq 1,\\
0, & 0 \leq \mu < \mu_0,
\end{array} \right.
\nonumber
\end{equation}

\noindent
where $B(T_*,\nu)$ is the intensity of blackbody radiation, $T_*$ is the temperature of the stellar photosphere, $\mu_0$ is the critical value of the parameter, $\mu_0 = cos\theta_0 = \sqrt{D^2 - R_*^2}/D$, $D$ is the distance from the stellar center to the cloud surface, and $R_*$ is the stellar radius. We assume that $H << D$ (plane geometry).

Each of the coefficients $\varepsilon(z,\mu,\nu)$ and $\kappa(z,\mu,\nu)$ is the sum of the emission or absorption coefficient in continuum and the emission or absorption coefficient in a spectral line, respectively:

\begin{equation}
\begin{array}{c}
\displaystyle
\varepsilon(z,\mu,\nu) = \varepsilon_c(\nu)+\frac{\displaystyle h\nu}{\displaystyle 4\pi}A_{ik} N n_i(z)\phi_{ik}(z,\mu,\nu), \nonumber\\ [10pt]
\displaystyle
\kappa(z,\mu,\nu) = \kappa_c(\nu) + \frac{\displaystyle \lambda^2}{\displaystyle 8\pi}A_{ik}N \left (\frac{\displaystyle g_i}{\displaystyle g_k}n_k(z)-n_i(z) \right) \phi_{ik}(z,\mu,\nu), \\[10pt]
\end{array}
\end{equation}

\noindent
where $\varepsilon_c(\nu)$ and $\kappa_c(\nu)$ are the emission and absorption coefficients of dust, $A_{ik}$ is the Einstein coefficient for spontaneous emission, $n_i(z)$ and $n_k(z)$ are the normalized populations of levels $i$ and $k$, $\sum_{j} n_j(z) =1$, $N$ is the particle number density, $g_i$ and $g_k$ are the statistical weights of the levels, $\lambda$ is the radiation wavelength, and $\phi_{ik}(z,\mu,\nu)$ is the normalized spectral line profile. In these formulas, it is implied that level $i$ lies above level $k$ in energy. The spectral profile of the emission and absorption coefficients in the laboratory frame of reference $\phi_{ik}(z,\mu,\nu)$ is

\begin{equation}
\displaystyle
\phi_{ik}(z,\mu,\nu)=\tilde{\phi}_{ik}(\nu-\mu\nu_{ik}v(z)/c),
\nonumber
\end{equation}

\noindent
where $\tilde{\phi}_{ik}(\nu)$ is the spectral line profile in the comoving frame of reference, $\nu_{ik}$ is the transition frequency, $v(z)$ is the gas velocity along the $z$ axis, here $v(z) = K_{V}z$, $K_V$ is a constant coefficient equal to the gas velocity gradient in the cloud. For the function $\tilde{\phi}_{ik}(\nu)$, we have

\begin{equation}
\tilde{\phi}_{ik}(\nu)=\frac{1}{\sqrt{\pi}\Delta\nu_{ik}} \text{exp} \left(-\left(\frac{\nu-\nu_{ik}}{\Delta\nu_{ik}} \right)^2 \right),
\nonumber
\end{equation}

\noindent
where $\Delta\nu_{ik}$ is the line profile width. The line profile width is determined by the spread in thermal velocities of molecules and turbulent velocities in the gas-dust cloud:

\begin{equation}
\Delta\nu_{ik}=\nu_{ik}\frac{v_D}{c}, \quad v_D^{2}=v_T^2+v_{\text{turb}}^2,
\nonumber
\end{equation}

\noindent
where $v_T = \sqrt{2kT_g/m}$ is the most probable thermal velocity of the molecules, $k$ is the Boltzmann constant, $T_g$ is the gas kinetic temperature, $m$ is the mass of the molecule, and $v_{\text{turb}}$ is the characteristic turbulent velocity in the cloud.

\subsection*{\textmd{\textit{The System of Statistical Equilibrium Equations for the Level Populations}}}

In the stationary case, the system of equations for the level populations is

\begin{equation}
\begin{array}{c}
\displaystyle
\sum_{k=1, \, k \ne i}^M \left( R_{ki}(z)+C_{ki} \right) n_k(z) - n_i(z)\sum_{k=1, \, k \ne i}^M \left( R_{ik}(z)+C_{ik} \right)=0, \quad i=1,...,M-1, \\
\displaystyle
\sum_{i=1}^M n_i(z)=1,
\end{array}
\end{equation}

\noindent
where $M$ is the total number of levels, $R_{ik}(z)$ is the rate coefficient for the transition from level $i$ to level $k$ through radiative processes, and $C_{ik}$ is the rate coefficient for the transition from one level to another through collisional processes. The rate coefficients for radiative transitions $R_{ik}(z)$ are

\begin{equation}
\begin{array}{c}
R_{ik}^{\downarrow}(z)=B_{ik}J_{ik}(z)+A_{ik}, \quad \varepsilon_i > \varepsilon_k, \\[10pt]
R_{ik}^{\uparrow}(z)=B_{ik}J_{ik}(z), \quad \varepsilon_i < \varepsilon_k,
\end{array}
\nonumber
\end{equation}

\noindent
where $\varepsilon_i$ and $\varepsilon_k$ are the level energies, $A_{ik}$ and $B_{ik}$ are the Einstein coefficients for spontaneous and stimulated emission, $J_{ik}(z)$ is the radiation intensity averaged over the direction and over the line profile:

\begin{equation}
J_{ik}(z)=\frac{1}{2} \int\limits_{-1}^{1}  d\mu \int\limits_{-\infty}^{\infty} d\nu \, \phi_{ik} (z,\mu,\nu) I(z,\mu,\nu),
\nonumber
\end{equation}

\noindent
where $I(z,\mu,\nu)$ is the solution of Eq. (2). The Einstein coefficients for spontaneous and stimulated emission $A_{ik}$, $B_{ik}$, and $B_{ki}$ are related by the relations

\begin{equation}
\displaystyle
A_{ik} \frac{\lambda^2}{2h\nu} = B_{ik} = \frac{g_k}{g_i}B_{ki}, \quad \varepsilon_i > \varepsilon_k.
\nonumber
\end{equation}

The transitions of H$_2$O molecules in collisions with He atoms and H$_2$ molecules are considered in the model. The following relation holds for the rate coefficients for the molecule's collisional excitation and deexcitation:

\begin{equation}
\displaystyle
C_{ik}^{\downarrow}=\frac{g_k}{g_i}C_{ki}^{\uparrow}\text{exp}(\frac{\varepsilon_i-\varepsilon_k}{kT_g}), \quad C_{ik}^{\downarrow}=\sum_l N_l r_{ik}^l(T_g),
\nonumber
\end{equation}

\noindent
where $N_l$ is the number density of collisional partners of type $l$, $r^l_{ik}(T_g)$ are the collisional rate coefficients. It is implied in this expression that $\varepsilon_i > \varepsilon_k$.

The emission from inverted transitions is disregarded in our calculations of the energy level populations for H$_2$O molecules.

\subsection*{\textmd{\textit{The Accelerated $\Lambda$-Iteration Method}}}

The statistical equilibrium equations for the level populations contain the rate coefficients for radiative transitions that are expressed in terms of the radiation intensity in the medium. In turn, the radiation intensity depends on the absorption and emission coefficients in molecular lines that are expressed in terms of the level populations. The radiative transfer equation in the medium (2) and the system of statistical equilibrium equations for the level populations (3) are solved self-consistently by the accelerated $\Lambda$-iteration method (Rybicki and Hummer 1991). Let us introduce the concept of an exact $\Lambda$-operator:

\begin{equation}
I(z,\mu,\nu)=\Lambda(z,\mu,\nu) \left[S^{\dagger} \right],
\nonumber
\end{equation}

\noindent
where the $\Lambda$-operator represents the set of all mathematical operations needed to calculate the radiation intensity based on the known level populations and source function $S^{\dagger}$. The acceleration of the iterative series is achieved through the "splitting" of the $\Lambda$-operator into an approximate operator and the remaining part:

\begin{equation}
\Lambda=\Lambda^*+(\Lambda-\Lambda^*).
\nonumber
\end{equation}

\noindent
The scheme for calculating the radiation intensity is modified as follows:

\begin{equation}
I(z,\mu,\nu)=\Lambda^*(z,\mu,\nu)[S]+(\Lambda(z,\mu,\nu) - \Lambda^*(z,\mu,\nu)) \left[S^{\dagger} \right].
\end{equation}

\noindent
The first term contains the unknowns at a given level population iteration step; the second term is calculated based on the level populations and source functions $S^\dagger$ derived in the preceding iteration. Equation (4) is substituted into the system of equations for the level populations (3) and "new" level populations are calculated. We chose the local operator from Rybicki and Hummer (1991) as the $\Lambda^*$-operator. The locality means that the level populations and source functions at depth $z$ are used in calculations of the first term in Eq. (4).

\subsection*{\textmd{\textit{Spectroscopic Data and Collisional Rate Coefficients}}}

In our calculations, we took into account 410 rotational levels of ortho-H$_2$O and 410 rotational levels of para-H$_2$O belonging to the five lowest vibrational levels of the ground electronic state of the molecule. The collisional and radiative transitions between the ortho- and para-spin-isomers of H$_2$O are forbidden in the dipole approximation. The first rotational levels of the excited vibrational levels have the following energies: 1594.7 cm$^{-1}$ for the (010) vibrational level, 3151.6 cm$^{-1}$ for (020), 3657.1 cm$^{-1}$ for (100), and 3755.9 cm$^{-1}$ for (001). The energy of the uppermost level among those under consideration is about 5000~cm$^{-1}$ or, in temperature units, 7200 K. The spectroscopic data for H$_2$O molecules were taken from the HITRAN 2008 database (Rothman et al. 2009). The energies of the rotational-vibrational transitions for the H$_2$O levels under consideration do not exceed 4300 cm$^{-1}$ (2.3 $\mu \text{m}$) and correspond to the infrared and the radio band. Note that the peak of the intensity of blackbody radiation $B(\nu)$ at a temperature of 2300 K is near 2.2 $\mu \text{m}$.

The rate coefficients for collisional transitions between H$_2$O levels in collisions of H$_2$O with H$_2$ were taken from Faure et al. (2007) and Faure and Josselin (2008). Faure and Josselin (2008) provided the rate coefficients for collisional transitions between rotational levels of the first five vibrational levels of H$_2$O. The collisional rate coefficients for transitions between H$_2$O levels in inelastic collisions of H$_2$O with He atoms were taken from Green et al. (1993). The data from Green et al. (1993) contain the collisional rate coefficients for the lowest 45 levels of ortho-H$_2$O and 45 levels of para-H$_2$O. The collisional rate coefficients for the transitions including higher H$_2$O levels were calculated by extrapolating the data from Green et al. (1993) using the algorithm proposed by Faure and Josselin (2008) for the collisional rate coefficients for H$_2$O and H$_2$. In our calculations, we used the relaxation rates of excited vibrational levels of H$_2$O for H$_2$O and He collisions from Kung and Center (1975).

\subsection*{\textmd{\textit{Optical Properties of Dust}}}

One of the main parameters defining the composition of dust particles in the circumstellar envelopes of late-type stars is the carbon-to-oxygen ratio in the stellar atmosphere (H$\ddot o$fner 2009). In the case where the relative oxygen abundance exceeds the carbon one, dust particles composed of metal oxides and silicates are predominantly formed. Numerical simulations of the stellar wind in the circumstellar envelopes of stars of this type (H$\ddot o$fner 2008; Sacuto et al. 2013) and polarimetric observations (Norris et al. 2012) point to the presence of dust grains with radii in the range 0.1-1 $\mu \text{m}$.

Here we use the complex dielectric function for dust from David and P$\acute e$gouri$\acute e$ (1995). In our calculations, the dust particle radius $a$ is assumed to be 0.3 $\mu \text{m}$. The cross sections for the absorption and scattering of radiation by dust particles and the mean scattering angle were calculated using the Mie scattering theory. We used the numerical code published in the monograph by Bohren and Huffman (1983) and modified by Draine \footnote[1]{http://code.google.com/p/scatterlib/wiki/Spheres} (2004). The wavelength dependence of the dust absorption coefficient in the long-wavelength infrared range ($\lambda >$ 150 $\mu \text{m}$) is extrapolated by a power law with an exponent of~-1.5 (David and P$\acute e$gouri$\acute e$ 1995). The dust emissivity is defined by the expression

\begin{equation}
\displaystyle
\varepsilon_c(\nu)=\kappa_c(\nu) \times \frac{2h\nu}{\lambda^2} \frac{1}{\text{exp} (h \nu /k T_d) - 1},
\nonumber
\end{equation}

\noindent
where $\kappa_c(\nu)$ is the dust absorption coefficient and $T_d$ is the dust temperature.

The fluxes of the gas and dust components in the stellar wind at distance $r$ from the star are

\begin{equation}
\begin{array}{c}
F_g(r)=\rho_g(r) v_g(r), \\[10pt]
F_d(r)=\rho_d(v_g (r)+v_{\text{drift}}(a,r)), \\[10pt]
\displaystyle
\rho_d = \frac{4\pi}{3} a^3 \rho_m N_d,
\end{array}
\nonumber
\end{equation}

\noindent
where $\rho_g$ is the gas density, $\rho_d$ is the cloud dust density, $v_g(r)$ is the gas velocity, $v_{\text{drift}}(a, r)$ is the drift velocity of the dust particles relative to the gas, $a$ is the particle radius, $N_d$ is the dust particle number density, and $\rho_m$ is the dust density, 3 g cm$^{-3}$. The dust particle drift velocity through the gas is determined by the balance between the stellar radiation pressure force and the force of friction from the gas. The expression for the dust particle drift velocity is (Kwok 1975; Decin et al. 2006)

\begin{equation}
\begin{array}{c}
\displaystyle
v^2_{\text{drift}}(a,r)=v_0^2(a,r) \left[ \sqrt{1+x^2(r)} - x(r) \right], \\[10pt]
\displaystyle
v_0^2(a,r) = \frac{\Omega_*(r)}{c \rho_g \pi a^2} \int \limits_0^{\infty} d\nu c_{\text{rp}} (a,\nu) B(T_*,\nu), \\[10pt]
\displaystyle
x(r)=\frac{1}{2} \frac{v_T^2}{v_0^2}, \quad v_T^2=\frac{27kT_g}{16 \overline{m}}, \\[10pt]
\displaystyle
\Omega_*(r)=2\pi \left(1 - \sqrt{1-R_*^2/r^2}\right),
\nonumber
\end{array}
\end{equation}

\noindent
where $\Omega_*(r)$ is the solid angle that cuts out the stellar disk on the celestial sphere at distance $r$ from the stellar center, $\overline{m}$ is the mean mass of the gas molecules and atoms, and $c_{\text{rp}}(a,\nu)$ is the cross section for the photon momentum transfer to a dust particle upon absorption and scattering. The parameter $c_{\text{rp}}(a,\nu)$ is calculated from the formula

\begin{equation}
c_{\text{rp}}(a,\nu)=c_{\text{abs}}(a,\nu)+c_{\text{sca}}(a,\nu)(1-\langle cos \theta \rangle),
\nonumber
\end{equation}

\noindent
where $c_{\text{abs}}(a, \nu)$ is the cross section for the absorption of a photon by a dust particle of radius $a$, $c_{\text{sca}}(a, \nu)$ is the cross section for the scattering of a photon by a dust grain, and $\langle cos \theta \rangle$ is the mean scattering angle. According to the estimates from Kwok (1975), the material evaporates from the particle surface at dust particle drift velocities of about 20 km s$^{-1}$ or higher. For a particle radius of 0.3 $\mu \text{m}$ and gas number densities $N_{H_2} > 10^7$ cm$^{-3}$, the dust particle drift velocity in the model under consideration does not exceed 20 km s$^{-1}$.

In our calculations, we took the dust-to-gas flux ratio in the stellar wind to be $f_d$ = 0.01, the mean dust-to-gas ratio in the interstellar medium (Whittet 2003). The dust-to-gas ratio in the cloud $\rho_d/\rho_g$ is determined from the equation $F_d(r)/F_g(r) = f_d$. The lower the gas density, the higher the dust particle drift velocity and, consequently, the dust-to-gas ratio in the cloud is smaller, other things being equal.

\subsection*{\textmd{\textit{Numerical Calculations}}}
The gas-dust cloud in our numerical model is broken down in $z$ coordinate into $N$ = 200 layers. The level populations within each layer are constant. The near-surface layers of the cloud have a thickness of 10$^{-6}H$ and the thickness of the succeeding layer deep into the cloud is larger than the thickness of the preceding one by a constant factor. The grids of values for the angle and the frequency are determined. The range of values for the parameter $\mu$ is [0;1]; the discretization step was chosen to be 0.1. In our calculations, we use the parameter $x = (\nu - \nu_{ik})/\Delta \nu_{ik}$ that characterizes the deviation of the radiation frequency from the transition frequency $\nu_{ik}$. The range of values for the parameter $x$ for each line was chosen to be [-5; 5]; the discretization step is 0.25.

At each iteration step, the radiative transfer equation in each spectral line is solved for each pair of values of the parameters $\mu$ and $x$. The radiative transfer equation (2) is written as a second-order differential equation (Feautrier 1964; Peraiah 2004). After the substitution of the differential operator by the ratios of finite differences, we set up a system of linear equations for the radiation intensities in the cloud layers. This system of equations is solved using the algorithm described by Rybicki and Hummer (1991). After averaging over the angle and the frequency, the radiation intensity and the local operator are substituted into the system of equations for the level populations (3). The number of equations in this system is $N \times M$, where $N$ is the number of cloud layers and $M$ is the number of levels. By solving the system of equations, we find the molecular level populations in each cloud layer that are used in the next iteration as input data. The initial level populations at the first iteration step are obtained by a method based on the photon escape probability.

An additional acceleration of the iterative series is achieved by applying the convergence optimization method proposed by Ng (1974). At some iteration step, the new vector of level populations is represented as a linear combination of the population vectors obtained in preceding iterations. The coefficients of the linear form are calculated by minimizing the vector of residuals. The convergence criterion for the iterative series is the condition on the maximum relative increment in level populations for two successive iterations, $\displaystyle \max_i |\Delta n_i/n_i| < 10^{-4}$. The statistical equilibrium equations for the level populations and the radiative transfer equation are self-consistently solved separately for the ortho-H$_2$O and para-H$_2$O molecules.

In our calculations, we used the algorithms for solving systems of linear equations published in the book by Press et al. (1997).

\section*{\textmd{The Thermal Balance of Gas and Dust in the Cloud}}
\subsection*{\textmd{\textit{The Thermal Balance of Dust}}}

The main dust heating mechanism in the inner part of the circumstellar envelope is the absorption of stellar radiation (Babkovskaia 2005). In our calculations of the dust heating rate, we use the approximation of an optically thin (in continuum) medium. The amount of heat absorbed by dust per unit gas volume per unit time is

\begin{equation}
q_d^{+}(r)=\Omega_*(r) \int \limits_0^{\infty} d \nu \, \kappa_c(\nu) B(T_*,\nu),
\nonumber
\end{equation}

\noindent
where $r$ is the distance from the stellar center to the test gas volume. The amount of heat lost by dust through its intrinsic radiation per unit gas volume per unit time is

\begin{equation}
q_d^-(T_d)=-4\pi \int \limits_0^{\infty} d \nu \, \varepsilon_c (\nu).
\nonumber
\end{equation}

\noindent
The dust heating or cooling through the processes of collisions of dust particles with gas atoms and molecules are insignificant compared to the radiative processes (Babkovskaia 2005). The expression for the thermal balance of dust is

\begin{equation}
q_d^+(r)+q_d^-(T_d)=0.
\nonumber
\end{equation}

\noindent
The dust temperature $T_d$ in the gas-dust cloud is determined from the solution of this equation. The dust temperature in the cloud in the model under consideration is 450 K.

\subsection*{\textmd{\textit{The Thermal Balance of Gas}}}

All the main processes of heating and heat removal from the gas-dust cloud should be taken into account to determine the gas temperature. We consider the gas heating due to the drift of dust particles through the gas, the heat exchange in the collisions of gas molecules and atoms with dust particles, the emission in molecular spectral lines, and the heat losses through adiabatic cloud expansion. The gas heating due to the photoelectric effect on dust particles and the gas heating through the interaction with cosmic ray particles are insignificant in the inner regions of the circumstellar envelopes around late-type stars (Babkovskaia 2005; Decin et al. 2006).

Radiation pressure is the driving force of the stellar wind in the circumstellar envelopes of AGB stars. Absorbing and scattering the stellar radiation, the dust particles experience a radiation pressure force. The dust transfers its momentum to the gas through the collisions of dust particles with gas atoms and molecules. The collisions of gas and dust particles also lead to gas heating. The amount of heat released per unit gas volume per unit time as a result of the dust drift through the gas is (Decin et al. 2006)

\begin{equation}
\displaystyle
q_{\text{drift}}(r) = \frac{1}{2} \rho_g \pi a^2 N_d v_{\text{drift}}^3 (a,r).
\nonumber
\end{equation}

The gas and dust in the cloud have different temperatures. As a result of the collisions of gas atoms and molecules with dust particles, thermal energy is transferred between them. The heat exchange rate between the gas and dust is (Burke and Hollenbach 1983)

\begin{equation}
\begin{array}{c}
\displaystyle
q_{\Delta T} = 2 \alpha k (T_d - T_g) \pi a^2 N_d \sum_i N_i \langle v_i \rangle, \\[10pt]
\displaystyle
\langle v_i \rangle = \sqrt{\frac{8kT_g}{\pi m_i}},
\end{array}
\nonumber
\end{equation}

\noindent
where $N_i$, $\langle v_i \rangle$, and $m_i$ are the number density, mean thermal velocity, and mass of the molecules or atoms of type $i$, respectively; $\alpha$ is the accommodation coefficient, $\alpha \approx$ 0.2. The parameter $q_{\Delta T}$ is positive when $T_d > T_g$ and negative in the opposite case.

The gas-dust cloud loses (gains) energy through the emission and absorption in molecular lines. The amount of energy lost or gained by a unit gas volume per unit time through the emission and absorption in H$_2$O lines is

\begin{equation}
\displaystyle
q_{H_2O}(r) = N_{H_2O} \sum_{i>k} h \nu_{ik} (C_{ik} n_i(r) - C_{ki} n_k(r)).
\nonumber
\end{equation}

\noindent
The parameter $q_{H_2O}$ calculated in this way is the sum of the rate of gas heating due to the absorption of the external radiation field and dust emission and the rate of cooling due to the emission in molecular lines. In our calculations, we take into account the cloud cooling (heating) due to the emission and absorption in ortho-H$_2$O and para-H$_2$O lines. According to the results by Decin et al. (2010), the gas cooling due to the emission in H$_2$O lines exceeds that for CO in the inner region of the envelope around IK Tau by more than an order of magnitude. In our calculations, we disregarded the gas cooling (heating) due to the emission and absorption in CO lines. 

The H$_2$ molecule is another coolant in the circumstellar envelopes of late-type stars. Because of their small Einstein coefficients, the H$_2$ lines are optically thin even in the inner dense regions of the stellar envelope (Decin et al. 2006). The amount of energy lost (gained) by a unit gas volume per unit time through the emission and absorption in H$_2$ lines can be estimated as

\begin{equation}
\displaystyle
q_{H_2}(r) = -N_{H_2} \sum_{i>k} A_{ik} h \nu_{ik} n_i + \frac{\Omega_*(r)}{4\pi} N_{H_2} \sum_{i>k} B_{ik} h \nu_{ik} B(T_*, \nu_{ik}) \left( \frac{g_i}{g_k} n_k - n_i \right),
\nonumber
\end{equation}

\noindent
where $n_i$ are the H$_2$ level populations and $B(T_*,\nu)$ is the intensity of the stellar radiation. In our calculations, we assume that the molecular hydrogen level populations have the Boltzmann distribution. We take into account 21 rotational levels of the molecule belonging to the ground and the first excited vibrational levels of H$_2$. The level energies and Einstein coefficients were taken from Dabrowski (1984) and Wolniewicz et al. (1998).

The expansion of the gas-dust envelope is responsible for the adiabatic heat losses. The amount of energy lost by a unit gas volume per unit time is

\begin{equation}
\displaystyle
q_{\text{ad}}(r)=-kNT_g \left( \frac{2v_g}{r} + \frac{dv_g}{dr} \right),
\nonumber
\end{equation}

\noindent
where $N$ is the gas particle number density. To estimate the mean velocity and mean velocity gradient of the gas, we used observational data on the velocity distribution of H$_2$O maser sources in the circumstellar envelope of IK Tau (Richards et al. 2011, 2012).

In our calculations of the gas temperature, we consider the gas heating and cooling rates averaged over the cloud, $\overline{q}_i = \frac{1}{H} \int_0^H dz \, q_i(z)$. The thermal balance equation for the gas is

\begin{equation}
\overline{q}_{\text{drift}} + \overline{q}_{\Delta T} + \overline{q}_{H_2O} + \overline{q}_{H_2} + \overline{q}_{\text{ad}} = 0.
\nonumber
\end{equation}

\noindent
The gas temperature is assumed to be independent of the coordinates. This assumption simplifies the calculations considerably. The gas temperature in the cloud is determined by the balance between gas heating and cooling rates, which depend on the H$_2$O level populations. At the same time, the H$_2$O level populations are determined by the rate coefficients for collisional and radiative transitions, which depend on the gas temperature in the cloud. To determine the gas temperature, the statistical equilibrium equations for the level populations and the thermal balance equation for the gas in the gas-dust cloud should be solved self-consistently. The equations are solved self-consistently by the method of dichotomy relative to the gas temperature.

The calculations were performed at the St. Petersburg branch of the Joint Supercomputer Center of the Russian Academy of Sciences.

\section*{\textmd{Results}}
\subsection*{\textmd{\textit{The Gas Heating and Cooling Rates}}}

In this section, we present the calculated gas cooling and heating rates in the cloud in various processes as a function of the gas temperature. The physical parameters of the cloud are given in the table. We took the number density of hydrogen molecules to be 10$^8$ cm$^{-3}$ and the total number density of H$_2$O molecules to be 10$^4$ cm$^{-3}$. For each gas temperature, we solve the statistical equilibrium equations for the level populations of ortho-H$_2$O and para-H$_2$O molecules and calculate the gas cooling and heating rates in the cloud.

\begin{figure}[t]
	\centering
	\includegraphics[width = 0.95\textwidth]{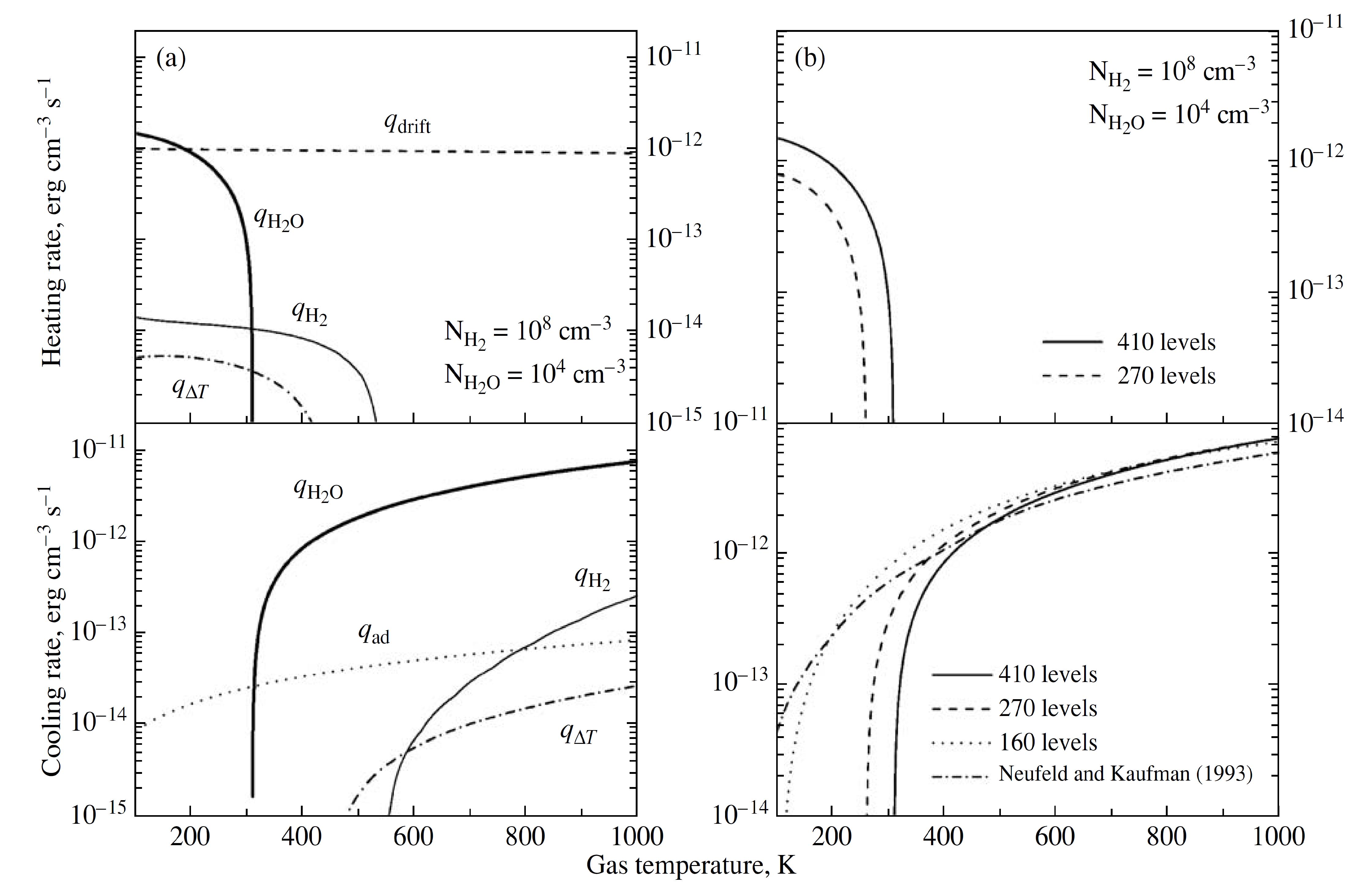}
		\caption{\small{(a) Gas heating and cooling rates in the gas-dust cloud in various processes. The rate of cooling (heating) due to the emission and absorption in H$_2$O lines (thick solid line); the rate of cooling (heating) due to the emission and absorption in H$_2$ lines (thin solid line); the rate of heating due to the drift of dust particles through the gas (dashed line); the rate of energy losses through adiabatic expansion (dotted line); the rate of heat exchange between the gas and dust particles (dash-dotted line). (b) The rates of gas cooling and heating due to the emission and absorption in H$_2$O lines. The results of our calculations in which 410 rotational levels belonging to the ground vibrational level and four excited vibrational levels of H$_2$O are taken into account (solid line); the results of our calculations in which 270 rotational levels belonging to the ground and the first excited vibrational levels of the molecule are taken into account (dashed line); the results of our calculations in which 160 rotational levels belonging to the ground vibrational level of the molecule are taken into account (dotted line). The gas cooling rate from Neufeld and Kaufman (1993) (dash-dotted line).}}
		\label{fig2}
\end{figure}

Figure 2a presents the calculated gas cooling and heating rates in the cloud. At low temperatures, the functions $\overline{q}_{H_2O}$ and $\overline{q}_{H_2}$ are positive, implying the gas heating due to the absorption of stellar radiation and dust emission in rotational-vibrational molecular lines. At high temperatures, the functions $\overline{q}_{H_2O}$ and $\overline{q}_{H_2}$ are negative. In this case, the gas cooling due to the emission in molecular lines dominates over the heating due to the absorption of radiation. The main processes responsible for the gas heating are the absorption of stellar radiation in H$_2$O molecular lines and the drift of dust particles through the gas. The emission in H$_2$O molecular lines is the main process responsible for the gas cooling at high temperatures.

Figure 2b presents the calculated rates of gas cooling and heating due to the emission and absorption in H$_2$O molecular lines. This figure presents the results of three calculations in which the following sets of levels are taken into account: the rotational levels belonging to the ground and four excited vibrational levels of the molecule (410 levels), the rotational levels belonging to the ground and the first excited vibrational levels of the molecule (270 levels), and the rotational levels belonging only to the ground vibrational level of H$_2$O (160 levels). The more vibrational and rotational levels are taken into account in the calculations, the more transitions are involved in the absorption of stellar radiation and the higher the calculated gas heating rate. The calculations in which the rotational levels of the excited vibrational H$_2$O levels are taken into account lead to higher equilibrium gas temperatures. In the range of high temperatures, including the molecule's excited vibrational levels does not lead to a significant increase in the gas cooling rate.

For comparison, Figure 2b presents the calculated cooling rates of the gas cloud due to the emission in H$_2$O lines from Neufeld and Kaufman (1993). The gas cooling rates averaged over the cloud are given for a static flat gas cloud. Neufeld and Kaufman (1993) disregarded the dust emission and the absorption of emission by dust and did not consider the external radiation field. The gas cooling rates obtained here and the results from Neufeld and Kaufman (1993) agree within 30$\%$ in the range of high temperatures.

\subsection*{\textmd{\textit{The Equilibrium Gas Temperature}}}

Figure 3 presents the calculated dependence of the equilibrium gas temperature on the molecular hydrogen number density $N_{H_2}$ and the total number density of water molecules $N_{H_2O}$ in the cloud. For each pair of $N_{H_2}$ and $N_{H_2O}$, the statistical equilibrium equations for the level populations of ortho-H$_2$O and para-H$_2$O molecules and the thermal balance equation for the gas are solved self-consistently. For H$_2$O number densities $N_{H_2O} \geq 10^4$ cm$^{-3}$, the gas temperature depends weakly on the parameters and lies within the range 370 K $< T_g <$ 450 K. For these H$_2$O number densities, the main processes determining the thermal balance of gas are the emission and absorption in rotational-vibrational H$_2$O lines. The equilibrium gas temperature is close to the value determined from the conditions of equality between the rate of gas cooling due to the emission in molecular lines and the rate of gas heating due to the absorption of stellar radiation in molecular lines. The gas heating due to the drift of dust particles and gas cooling (heating) due to the absorption and emission in the H$_2$ lines play an important role in the thermal balance at low H$_2$O number densities. The gas temperature increases with decreasing H$_2$O number density and decreasing H$_2$ number density.

\begin{figure}[t]
	\centering
	\includegraphics[width = 0.6\textwidth]{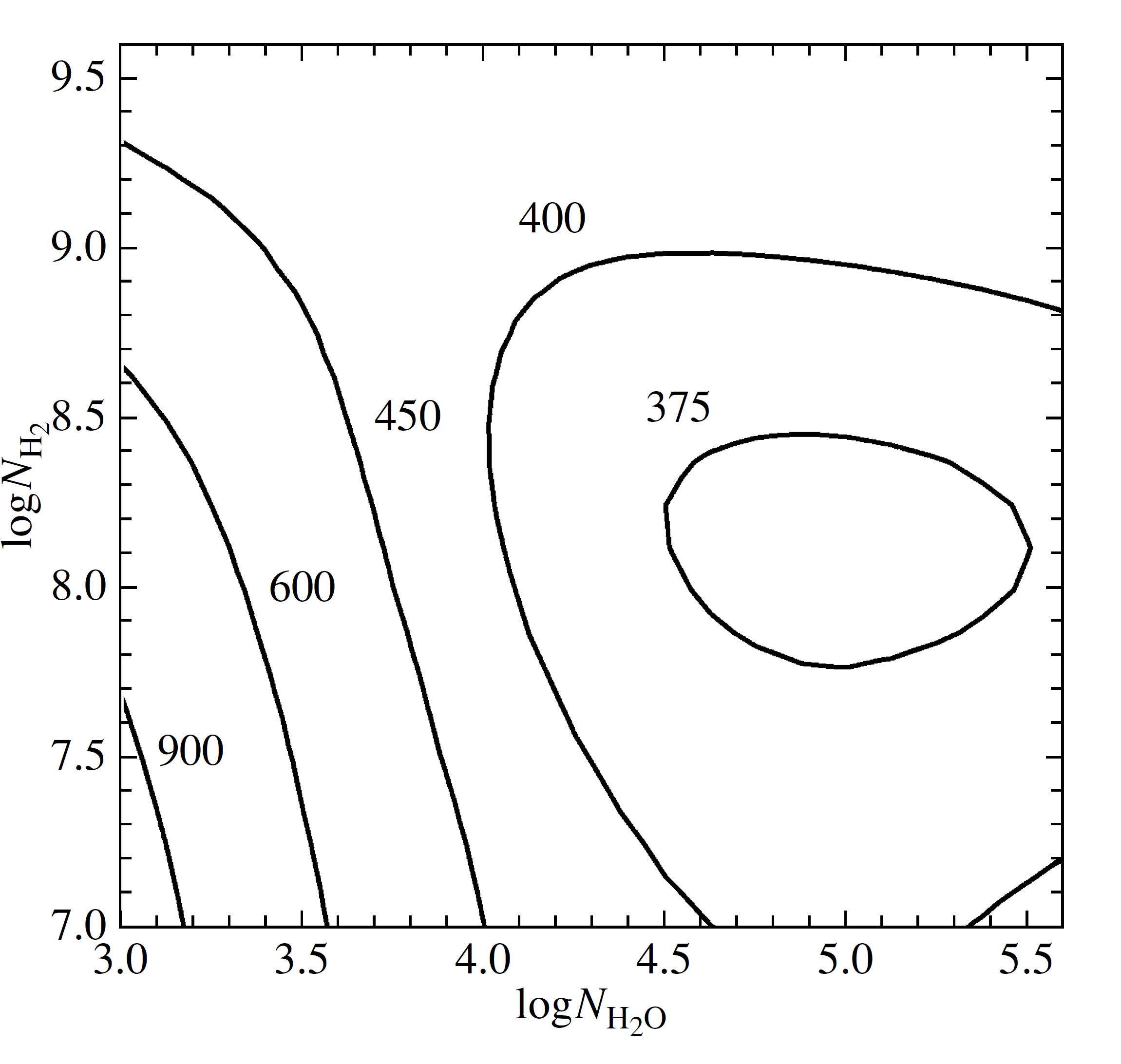}
		\caption{\small{Equilibrium gas temperature in the cloud as a function of the number density of hydrogen molecules $N_{H_2}$ and the total number density of water molecules $N_{H_2O}$. The contours of equal gas temperature are shown; the gas temperature in kelvins is indicated near each curve.}}
		\label{fig3}
\end{figure}

\subsection*{\textmd{\textit{The Gain in the 22.2-GHz Maser Line}}}

In the presence of an inverted level population, the line absorption coefficient $\kappa_{ij}$ is negative. Let us define the gain $\gamma_{ij} = -\kappa_{ij}$. The expression for $\gamma_{ij}$ at the line center in a direction along the cloud plane is

\begin{equation}
\displaystyle
\gamma_{ij}(z)=\frac{\lambda^2 A_{ik} N}{8 \pi \sqrt{\pi} \Delta \nu_{ij}} \left(n_i(z)-\frac{g_i}{g_k}n_k(z) \right),
\nonumber
\end{equation}

\noindent
where $N$ is the number density of molecules (ortho-H$_2$O or para-H$_2$O). When calculating the width of the spectral gain profile $\Delta \nu_{ik}$ in the 22.2-GHz ortho-H$_2$O line, we should take into account the additional profile broadening due to the hyperfine splitting (Varshalovich et al. 2006; Nesterenok and Varshalovich 2011):

\begin{equation}
\Delta \nu_{ik} = \nu_{ik} \frac{v}{c}, \quad v^2 \approx (v_T+0.5 \text{ km s}^{-1})^2 +v_{\text{turb}}^2,
\nonumber
\end{equation}

\noindent
where $\nu_{ik}$ is the mean transition frequency, $v_T = \sqrt{2kT_g/m}$. Figure 4 presents the calculated mean gain $\overline{\gamma}$ in the 22.2-GHz maser line, $\overline{\gamma} = \frac{1}{H} \int_{0}^{H} dz \, \gamma(z)$. The gain is at a maximum in the range of low H$_2$ number densities and high number densities of ortho-H$_2$O molecules.

The mean brightness temperature of the 22.2-GHz maser sources in the circumstellar envelope of IK Tau varies within the range 10$^8$-10$^{10}$ K (Richards et al. 2011). Thus, the optical depth in the maser line is about 15-20. When this parameter is estimated, the signal level excitation temperature is assumed to be $\sim$10 K and the background radio emission is neglected. The length of the amplification region along the line of sight probably does not exceed or is comparable to the distance from the cloud to the star $D$. Hence the gain in the maser line is estimated to be $>5 \times 10^{-14}$ cm$^{-1}$. According to our calculations, these gains are reached at number densities of hydrogen molecules in the range 10$^7$-10$^8$ cm$^{-3}$ and high H$_2$O abundances (relative to the hydrogen atoms), more than 10$^{-4}$. The signal level excitation temperature for this range of physical parameters varies between -10 and -1 K.

\begin{figure}[t]
	\centering
	\includegraphics[width = 0.6\textwidth]{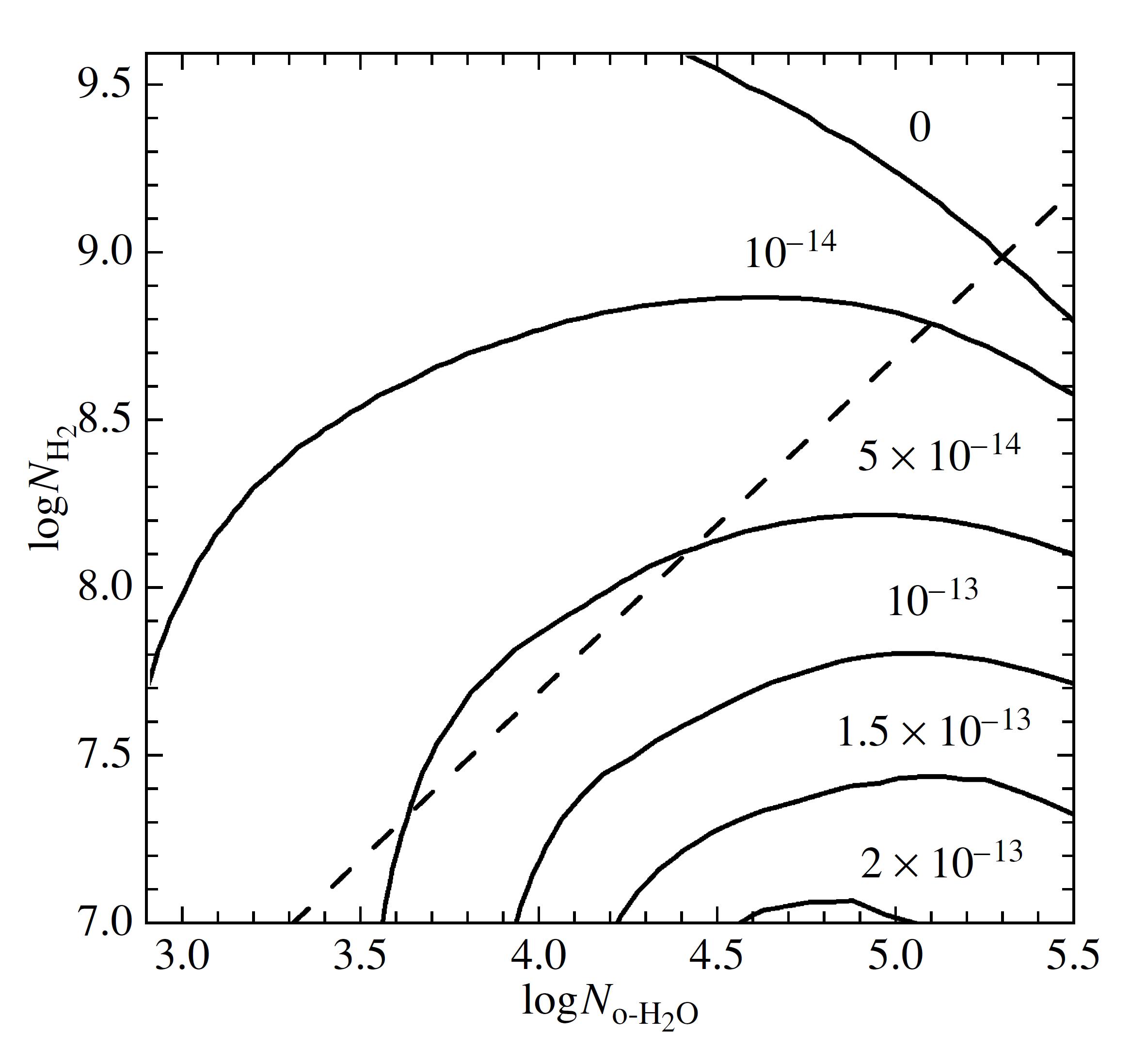}
		\caption{\small{Mean gain in the 22.2-GHz maser line as a function of the number density of hydrogen molecules $N_{H_2}$ and the number density of ortho-H$_2$O molecules $N_{\text{o-}H_2O}$. The contours of equal gain are shown; the gain in cm$^{-1}$ is indicated near each curve. The inclined dashed line corresponds to the points for which the relative ortho-H$_2$O abundance $N_{\text{o-}H_2O}/2N_{H_2}$ is 10$^{-4}$.}}
		\label{fig4}
\end{figure}

\section*{\textmd{Discussion}}
According to the results obtained here, molecular hydrogen number densities of 10$^7$-10$^8$ cm$^{-3}$ are needed to explain the observed intensity of maser sources in the circumstellar envelope of the AGB star IK Tau. These molecular hydrogen number densities are comparable in order of magnitude to the mean gas number density in the stellar wind (see Eq. (1)). Our results contradict the view (Richards et al. 1998, 1999, 2012; Bains et al. 2003; Murakawa et al. 2003) that the H$_2$O maser emission in the circumstellar envelopes of late-type stars originates in gas-dust clouds with a gas density of $\sim10^9$ cm$^{-3}$. According to our calculations, the gain in the 22.2-GHz maser line is less than 10$^{-14}$ cm$^{-1}$ for such densities. Note that Babkovskaia and Poutanen (2006) showed that the pumping of 22.2-GHz maser in the circumstellar envelopes is more efficient at H$_2$ number densities 10$^7$-10$^8$ cm$^{-3}$ and high H$_2$O relative abundances, which agrees with our results.

The collisional excitation of H$_2$O molecules to higher lying levels followed by the radiative deexcitation of these levels is the main H$_2$O maser pumping mechanism (Yates et al. 1997). In this case, for each photon of the emission being amplified there must be one or more infrared "sink" photons, which must either be absorbed by cold dust or escape from the resonance region to complete the pumping cycle. The collisional H$_2$O maser pumping is most efficient under conditions when the dust temperature is much lower than the gas temperature (Bolgova et al. 1977; Chandra et al. 1984; Yates et al. 1997; Babkovskaia and Poutanen 2004). The higher the dust temperature in the gas-dust cloud, the lower the gain. For example, according to the results by Yates et al. (1997), an increase in the dust temperature in the gas-dust cloud from 100 to 300 K can cause a tenfold decrease in the gain in the 22.2-GHz maser line. According to our calculations, the gas and dust temperatures in the cloud have close values for a wide range of gas and H$_2$O number densities. This is responsible for the relatively low gains in the 22.2-GHz maser line obtained in our calculations.

In our calculations, we used new data on the rate coefficients for collisional transitions from Faure et al. (2007) and Faure and Josselin (2008). The collisional rate coefficients for H$_2$O transitions in inelastic collisions of H$_2$O with H$_2$ published by Faure et al. (2007) exceed those obtained using the rate coefficients from Green et al. (1993) by a factor of 1-3. Consequently, the calculations in which the collisional rate coefficients from Faure et al. (2007) are used lead to lower admissible molecular hydrogen number densities in maser sources (see Section 3.3 in Faure et al. 2007).

Decin et al. (2010) provide an estimate of the relative H$_2$O abundance in the circumstellar envelope of IK Tau $N_{H_2O}/2N_{H_2}$ = $3.3 \times 10^{-5}$ and an ortho-to-para-H$_2$O ratio of 3. This estimate was obtained by analyzing the observational data from the Herschel satellite and the numerical simulations of physical processes in the stellar envelope. Here, we showed that a high relative H$_2$O abundance, $N_{H_2O}/2N_{H_2} > 10^{-4}$, is needed to explain the observed intensities of H$_2$O maser sources. The relative H$_2$O abundance in maser sources can differ significantly from the mean value of this parameter in the stellar wind.

H$_2$O molecules are formed in the inner hot regions of a red giant's gas-dust envelope. Molecules can be formed both under thermodynamic chemical equilibrium conditions and under chemically nonequilibrium conditions triggered by the passage of shocks through the gas (Cherchneff 2006). A significant fraction of oxygen can be contained in dust grains and CO molecules. Consequently, the relative H$_2$O abundance in the stellar wind must be lower than the relative oxygen abundance in the stellar photosphere in the absence of any other H$_2$O sources in the stellar envelope.

Cometary bodies can be H$_2$O sources in the circumstellar envelopes of late-type stars (Stern et al. 1990). The idea of the evaporation of cometary bodies was considered by Melnick et al. (2001), Justtanont et al. (2005), and Maercker et al. (2008) to explain the observed high H$_2$O abundance in the circumstellar envelopes of AGB stars. The relative H$_2$O abundance in the gas-dust cloud formed through the evaporation of a cometary body can be high. Investigating the parameters of the HDO emission in the circumstellar envelopes of AGB stars with a high angular resolution would allow the contribution from cometary bodies to the production of water in H$_2$O maser sources to be determined (Maercker et al. 2008).

\section*{\textmd{Conclusions}}
We investigated the physical conditions for the generation of 22.2-GHz H$_2$O maser emission in gas-dust clouds in the circumstellar envelopes of AGB stars. We took into account the main processes of heating and heat removal from the gas-dust cloud. Including the rotational levels belonging to the excited vibrational levels of H$_2$O leads to a significant increase in the calculated gas heating rate. The gas heating due to the absorption of stellar radiation in H$_2$O molecular lines and the gas cooling due to the emission in molecular lines are the main processes that determine the thermal balance of gas at H$_2$O number densities $\geq 10^4 \text{ cm}^{-3}$. We found that gas number densities of 10$^7$-10$^8$ cm$^{-3}$ and high H$_2$O relative abundances, $N_{H_2O}/2N_{H_2} > 10^{-4}$, are needed for the generation of intense maser emission.

\section*{\textmd{Acknoledgments}}
I wish to thank D.A. Varshalovich for a helpful discussion of the paper. This work was supported by the Russian Foundation for Basic Research (project no. 11-02-01018a), the Program of the President of Russia for Support of Leading Scientific Schools (project no. NSh-4035.2012.2), the Ministry of Education and Science of the Russian Federation (contract no. 8409, 2012), the Research Program OFN-17, the Division of Physics of the Russian Academy of Sciences.

\section*{\textmd{References}}
1. Alcock C. and R. R. Ross, Astrophys. J. 305, 837 (1986).

\noindent
2. Babkovskaia N., PhD Thesis (2005).

\noindent
3. Babkovskaia N. and J. Poutanen, Astron. Astrophys. 418, 117 (2004).

\noindent
4. Babkovskaia N. and J. Poutanen, Astron. Astrophys. 447, 949 (2006).

\noindent
5. Bains I., R. J. Cohen, A. Louridas, et al., Mon. Not. R. Astron. Soc. 342, 8 (2003).

\noindent
6. Bohren C. F. and D. R. Huffman, Absorption and Scattering of Light by Small Particles (Wiley, New York, 1983).

\noindent
7. Bolgova G. T., V. S. Strelnitskii, and I. K. Shmeld, Sov. Astron. 21, 468 (1977).

\noindent
8. Burke J. R. and D. J. Hollenbach, Astrophys. J. 265, 223 (1983).

\noindent
9. Chandra S., W. H. Kegel, D. A. Varshalovich, et al., Astron. Astrophys. 140, 295 (1984).

\noindent
10. Cherchneff  I., Astron. Astrophys. 456, 1001 (2006).

\noindent
11. Cooke B. and M. Elitzur, Astrophys. J. 295, 175 (1985).

\noindent
12. Dabrowski I., Canad. J. Phys. 62, 1639 (1984).

\noindent
13. David P. and B. P$\acute{e}$gouri$\acute{e}$, Astron. Astrophys. 293, 833 (1995).

\noindent
14. Decin L., Adv. Space Res. 50, 843 (2012).

\noindent
15. Decin L., S. Hony, A. de Koter, et al., Astron. Astrophys. 456, 549 (2006).

\noindent
16. Decin L., K. Justtanont, E. De Beck, et al., Astron. Astrophys. 521, L4 (2010).

\noindent
17. Deguchi S., Publ. Astron. Soc. Jpn. 29, 669 (1977).

\noindent
18. Draine B. T., The Cold Universe, Saas-Fee Advanced Course, Vol. 32, Ed. by A. W. Blain, F. Combes, B. T. Draine, D. Pfenniger, and Y. Revaz (Springer, Berlin, 2004), p. 213.

\noindent
19. Faure A. and E. Josselin, Astron. Astrophys. 492, 257 (2008).

\noindent
20. Faure A., N. Crimier, C. Ceccarelli, et al., Astron. Astrophys. 472, 1029 (2007).

\noindent
21. Feautrier P., Compt. Rend. Acad. Sci. Paris 258, 3189 (1964).

\noindent
22. Glassgold A. E. and P. J. Huggins, Mon. Not. R. Astron. Soc. 203, 517 (1983).

\noindent
23. Gonzalez Delgado D., H. Olofsson, F. Kerschbaum, et al., Astron. Astrophys. 411, 123 (2003).

\noindent
24. Green S., S. Maluendes, and A. D. McLean, Astrophys. J. Suppl. Ser. 85, 181 (1993).

\noindent
25. H$\ddot{o}$fner S., Astron. Astrophys. 491, L1 (2008).

\noindent
26. H$\ddot{o}$fner S., ASP Conf. Ser. 414, 3 (2009).

\noindent
27. Humphreys E. M. L., J. A. Yates, M. D. Gray, et al., Astron. Astrophys. 379, 501 (2001).

\noindent
28. Justtanont K., P. Bergman, B. Larsson, et al., Astron. Astrophys. 439, 627 (2005).

\noindent
29. Kung R. T. V. and R. E. Center, J. Chem. Phys. 62, 2187 (1975).

\noindent
30. Kwok S., Astrophys. J. 198, 583 (1975).

\noindent
31. Maercker M., F. L. Schoier, H. Olofsson, et al. Astron. Astrophys. 479, 779 (2008).

\noindent
32. Melnick G. J., D. A. Neufeld, K. E. S. Ford, et al., Nature 412, 160 (2001).

\noindent
33. Monnier J. D., R. Millan-Gabet, P. G. Tuthill, et al., Astrophys. J. 605, 436 (2004).

\noindent
34. Murakawa K., J. A. Yates, A. M. S. Richards, et al., Mon. Not. R. Astron. Soc. 344, 1 (2003).

\noindent
35. Neri R., C. Kahane, R. Lucas, et al., Astron. Astrophys. Suppl. Ser. 130, 1 (1998).

\noindent
36. Nesterenok A. V. and D. A. Varshalovich, Astron. Lett. 37, 456 (2011).

\noindent
37. Neufeld D. A. and M. J. Kaufman, Astrophys. J. 418, 263 (1993).

\noindent
38. Ng K.-C., J. Chem. Phys. 61, 2680 (1974).

\noindent
39. Norris B. R. M., P. G. Tuthill, M. J. Ireland, et al., Nature 484, 220 (2012).

\noindent
40. Olofsson H., M. Lindqvist, L-A. Nyman, et al., Astron. Astrophys. 329, 1059 (1998).

\noindent
41. Peraiah A., An Introduction to Radiative Transfer (Cambridge Univ. Press, Cambridge, UK, 2004).

\noindent
42. Press W. H., S. A. Teukolsky, W. T. Vetterling, et al., Numerical Recipes in C. The Art of Scientific Computing (Cambridge Univ. Press, Cambridge, 1997).

\noindent
43. Richards A. M. S., J. A. Yates, and R. J. Cohen, Mon. Not. R. Astron. Soc. 299, 319 (1998).

\noindent
44. Richards A. M. S., J. A. Yates, and R. J. Cohen, Mon. Not. R. Astron. Soc. 306, 954 (1999).

\noindent
45. Richards A. M. S., M. Elitzur, and J. A. Yates, Astron. Astroph. 525, A56 (2011).

\noindent
46. Richards A. M. S., S. Etoka, M. D. Gray, et al., Astron. Astroph. 546, A16 (2012).

\noindent
47. Rothman L. S., I. E. Gordon, A. Barbe, et al., J. Quant. Spectrosc. Rad. Transfer 110, 533 (2009).

\noindent
48. Rybicki G. B. and D. G. Hummer, Astron. Astrophys. 245, 171 (1991).

\noindent
49. Sacuto S., S. Ramstedt, S. H$\ddot{o}$fner, et al., Astron. Astrophys. 551, A72 (2013).

\noindent
50. Stern S. A., J. M. Schull, and J. C. Brandt, Nature 345, 305 (1990).

\noindent
51. Varshalovich D. A., A. V. Ivanchik, and N. S. Babkovskaya, Astron. Lett. 32, 29 (2006).

\noindent
52. Whittet D. C. B., Dust in the Galactic Environment (IOP, Bristol, 2003).

\noindent
53. Wolniewicz L., I. Simbotin, and A. Dalgarno, Astrophys. J. Suppl. Ser. 115, 293 (1998).

\noindent
54. Yates J. A., D. Field, and M. D. Gray, Mon. Not. R. Astron. Soc. 285, 303 (1997).

\bigskip
\noindent
Translated by V. Astakhov
\end{document}